\documentclass[%
 reprint,
superscriptaddress,
amsmath,amssymb,
twocolumn,
aps,
pra,
]{revtex4-2}

\usepackage{graphicx}
\usepackage{dcolumn}
\usepackage{bm}
\usepackage{amsthm}
\RequirePackage[usenames,dvipsnames]{xcolor}
\usepackage[
  pdfborder={0 0 0},
  colorlinks = true,
  citecolor  = RoyalBlue,
  linkcolor  = RoyalBlue,
  urlcolor   = RoyalBlue,
  unicode,
  ]{hyperref}   
\usepackage{braket}
\usepackage{spverbatim}
\usepackage{mathtools}
\usepackage{physics}
\usepackage{bm}
\usepackage{float}
\usepackage{comment}
\usepackage[capitalise]{cleveref}
\usepackage{booktabs}
\usepackage{tikz}

\usetikzlibrary{quantikz2}
\usepackage[linesnumbered,ruled,vlined]{algorithm2e}

\SetCommentSty{mycommfont}

\newcommand{\Dq}[1]{\hat{D}_q(#1)}

\newcommand{\Q}{\hat{\bm{q}}}

\newcommand{\CPHASE}{\hat{\bm{C}}^{\mathrm{CV}}_Z}
\newcommand{\CZ}{\hat{\bm{C}}_Z}

\newcommand{\CPHASEdag}{\hat{\bm{C}}^{\mathrm{CV}\dagger}_Z}
\newcommand{\CZdag}{\hat{\bm{C}}^\dagger_Z}
\newcommand{\RZ}{\hat{\bm{R}}_Z} 
\newcommand{\CD}{\hat{\bm{C}}_D}

\definecolor{bluemunsell}{rgb}{0.0, 0.5, 0.69}
\definecolor{ao}{rgb}{0.0, 0.5, 0.0}

\tikzset{
 noisy/.style={starburst,fill=yellow,draw=red,line
    width=1pt,inner xsep=-3pt,inner ysep=-3pt, starburst point height=9pt}
}

\newcommand{\SFU}{Department of Physics,
Simon Fraser University, Burnaby, British Columbia V5A 1S6, Canada}

\renewcommand{\selectlanguage}[1]{}
\begin{document}

\title{Downloading many-qubit entanglement from continuous-variable cluster states}

\author{Zhihua Han} \email{zhi\_han@sfu.ca} \affiliation{\SFU}

\author{Hoi-Kwan Lau} \email{kero\_lau@sfu.ca} \affiliation{\SFU}

\date{\today}
\begin{abstract}

Many-body entanglement is an essential resource for many quantum technologies, but its scalable generation has been challenging on qubit platforms. However, the generation of continuous-variable (CV) entanglement can be extremely efficient, but its utility is rather limited. In this work, we propose a scheme to combine the best of both qubit and CV approaches: a systematic method to download useful many-qubit entanglement from the efficiently generated CV cluster states. Our protocol is based on one-bit teleportation of the qubit correlation encoded in the displaced Gottesman-Kitaev-Preskill basis. To characterize the practical performance of our scheme, we develop an equivalent circuit to map dominant CV errors to single-qubit preparation errors. Particularly, we relate finite squeezing error to qubit erasure, and show that only 5.4~dB squeezing is sufficient to implement robust qubit memory or quantum computation (QC), and 11.9~dB for fault-tolerant QC. Our protocol can be implemented with the operations that are common in many bosonic platforms.

\end{abstract}

\maketitle


\noindent{\it Introduction.---} Many-body entanglement is essential to achieve the advantages of many quantum technologies, such as quantum sensing \cite{2017RvMP...89c5002D} and quantum computation (QC) \cite{10.1098/rspa.2002.1097}. One particular promising class of many-body entangled states is qubit cluster states (CS), which can be converted to any qubit entangled states by single-qubit measurement \cite{raussendorf_one-way_2001, raussendorf_measurement-based_2003}. Since discovery, qubit CS has become the backbone of numerous applications, including fault-tolerant (FT) QC \cite{bolt_foliated_2016}, all-photonic quantum repeaters \cite{azuma_all-photonic_2015, 10.1103/physreva.85.062326}, blind quantum computation \cite{Broadbent:2009uj}, and many more \cite{friis_flexible_2017}. Experimentally, qubit CS are usually generated in a ``bottom-up" approach in which large clusters are formed by sewing together smaller clusters. However, this approach has apparently encountered bottlenecks; experiments have demonstrated qubit CS with at most 51 qubits to date \cite{cao_generation_2023, thomas_efficient_2022, 10.1103/physrevlett.111.210501}.

On the other hand, the continuous variable (CV) extension of CS can be generated extremely efficiently \cite{menicucci_universal_2006}. Since Gaussian gates are readily implementable in most bosonic platforms, CVCS have been generated among millions of optical modes \cite{yoshikawa_invited_2016, asavanant_generation_2019, larsen_deterministic_2019,roh_generation_2025}. Their realization is recently extended to superconductor \cite{lingua_continuous-variable_2024, jolin_multipartite_2023, hernandez_control_2024} and trapped atomic systems \cite{cooper_graph_2024}, either with coherent gates or engineered dissipation \cite{zippilli_dissipative_2021}. Despite the generation efficiency, the utility of CVCS is rather limited. First, although ideal CVCS are universal resources for CV quantum information processing \cite{menicucci_universal_2006}, much fewer applications are developed for CV compared to the qubit counterpart \cite{braunstein_quantum_2005, weedbrook_gaussian_2012}. Second, realistic CVCS must incorporate only finite energy, but this introduces intrinsic noise to the information being processed \cite{ohliger_limitations_2010}. 

Here, we propose a scheme to combine the advantages of both the qubit and CV regimes: we introduce a ``top-down" method to generate versatile qubit CS by downloading the many-body entanglement from efficiently generated CVCS. Our method is based on teleporting the correlation embedded in CVCS to auxiliary physical qubits, and can be implemented with operations commonly found in many hybrid quantum systems \cite{Andersen:2015dp, kurizki_quantum_2015, liu_hybrid_2024}. To analyze the practical performance of our scheme, we present an equivalent circuit model to map the prominent CV imperfections to single-qubit preparation errors. In particular, we relate finite squeezing error with qubit erasure, and deduce a squeezing threshold of $11.9$ dB for FTQC, and only $5.4$ dB is sufficient for implementing standard QC or robust quantum memory.

\begin{figure*}
    \centering
    \includegraphics[width=\textwidth]{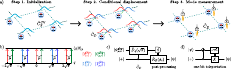}
    \caption{ a) Three-step protocol to download a qubit cluster state from a CV cluster state. 
    b) Flat wavefunction of $\ket{0}_p$ can be viewed as an equal superposition of displaced GKP basis states.
    c) Circuit corresponds to Steps 2 and 3 of the protocol. The initial qumode state is assumed a general displaced GKP qubit $\ket{\psi^\textrm{gkp}_{\mu,\nu}}$. d) Standard one-bit teleportation circuit.
}
    \label{fig:combined1}
\end{figure*}

\noindent{\it Entanglement downloading protocol.---} An $N$-qubit CS $\ket{G}$ is defined as a state generated by first initializing all qubits as $\ket{+} = \frac{1}{\sqrt{2}}(\ket{0} + \ket{1})$, and then applying controlled-Z (CZ) gates between the qubits that are specified by the edges of a graph $G = (N, E)$, i.e. 
$\ket{G} \equiv \prod_{(i,j) \in E}\hat{C}_{Z_{ij}}\ket{+}^{\otimes {N}}$ \cite{briegel_persistent_2001}. For later reference, we state the CZ gate definition as $\hat{C}_{Z_{ij}}\ket{lk}_{ij} = (-1)^{lk}\ket{lk}_{ij}$ for qubits $i$ and $j$ in the computational basis $l, k \in \{0, 1\}$.
Likewise, an ideal $N$-mode CVCS $\ket*{G^\mathrm{CV}} \equiv \prod_{(i,j) \in E}\hat{C}^{\mathrm{CV}}_{Z_{ij}}\ket{0}_p^{\otimes N}$ is generated by initializing all qumodes as $0$ $p$-quadrature eigenstates $\ket{0}_p$, and then CV controlled-phase (CPHASE) gates $\hat{C}_{Z_{ij}}^{\mathrm{CV}} \equiv e^{i\hat{q}_i \hat{q}_j}$ are applied between qumodes according to the graph $G$ \cite{menicucci_universal_2006}. 
Here, $\hat{q}$, $\hat{p}$ quadratures are related to the annihilation operator as $\hat{a}\equiv (\hat{q}+i\hat{p})/\sqrt{2}$, where $[\hat{a},\hat{a}^\dag]=1$. Their corresponding eigenstates are given by $\hat{q} \ket{s}_q = s \ket{s}_q$ and $\hat{p} \ket{t}_p = t \ket{t}_p$. For simplicity, 
we define the compact notation $\CZ \equiv \prod_{(i,j) \in E}\hat{C}_{Z_{ij}}$ and $\CPHASE \equiv \prod_{(i,j) \in E}\hat{C}^{\mathrm{CV}}_{Z_{ij}}$ to represent all qubit CZ and CV CPHASE gates required to generate the CS.

Our entanglement downloading protocol involves three steps, as shown in Fig. \ref{fig:combined1}(a):

\noindent{\textbf{Step 1.}} Prepare a CVCS $\ket*{G^\mathrm{CV}}$. Each qumode is associated with an auxiliary qubit initialized as $\ket{+}$.  

\noindent{\textbf{Step 2.}} Conditional displacement (CD) gates are applied between every qumodes and their associated qubit ancilla. For clarity, we represent the collection of all CD gates as $\CD \equiv \prod_{i = 1}^N \hat{C}_{D_i}$, where $\hat{C}_D \equiv \hat{I} \otimes \ketbra{0}{0} + e^{-i \sqrt{\pi} \hat{p}} \otimes \ketbra{1}{1}$.

\noindent{\textbf{Step 3.}} Every qumode is measured in $q$-quadrature. Upon obtaining the measurement outcomes $\bm{q} \equiv (q_1~q_2\ldots q_N)^T$, corrective phase shift is applied to all qubits, i.e. $\RZ \equiv \prod_{i = 1}^N \hat{R}_{Z_i}(\phi_i)$, where the phase-shift operator on $i$th qubit is $\hat{R}_{Z_i}(\phi_i) = e^{-i\hat{Z}_i \phi_i/2}$. The phases $\bm{\phi} \equiv(\phi_1~\phi_2\ldots \phi_N)^T$ are related to the measurement outcomes as $\bm{\phi} \equiv \sqrt{\pi}\mathbf{A}\bm{q}$, where $\mathbf{A}$ is the adjacency matrix of graph $G$, i.e. $A_{ij} = 1$ if $(i, j) \in E$ and $0$ otherwise. 

After these steps, we find that the auxiliary qubits will inherit a qubit CS with the same graph as the initial CVCS, i.e.
\begin{equation}
\RZ \bra{\bm{q}}\CD \ket*{G^\mathrm{CV}} \ket{+}^{\otimes N} \propto \ket{G}~,
\end{equation}
where $\ket{\bm{q}}\equiv \ket{q_1}_q\otimes \ldots \ket{q_N}_q$. This is our first main result.

\noindent{\textit{Protocol as hybrid teleportation.---}  The principle behind our scheme can be understood from three facts. First, every bosonic state can be expressed as a superposition of displaced Gottesman-Kitaev-Preskill (GKP) states \cite{gottesman_encoding_2001, glancy_error_2006}, where the basis states are $\ket*{0^\textrm{gkp}_{\mu,\nu}}\equiv \hat{D}_q(\mu)\hat{D}_p(\nu)\sum_m \ket{2m\sqrt{\pi}}_q$ and $\ket*{1^\textrm{gkp}_{\mu,\nu}}\equiv \hat{D}_q(\mu)\hat{D}_p(\nu)\sum_m \ket{(2m+1)\sqrt{\pi}}_q$; the displacement operators are $\hat{D}_q(\mu) = e^{-i\mu\hat{p}}$ and $\hat{D}_p(\nu) = e^{i\nu\hat{q}}$, and $\mu, \nu \in [-\sqrt{\pi}/2, \sqrt{\pi}/2)$ are the modular displacements in $q$ and $p$ respectively. For $\ket{0}_p$, as illustrated in Fig.~\ref{fig:combined1}(b), its wavefunction is real and flat in $q$ quadrature, so it can be represented as an equal superposition of all displaced GKP states that have no phase difference between their $q$-eigenstate components, i.e.
\begin{equation}
    \ket{0}_p \propto \int_{-\sqrt{\pi}/2}^{\sqrt{\pi}/2} d\mu  \ket*{0_{\mu, 0}^\textrm{gkp}}+ \ket*{1_{\mu, 0}^\textrm{gkp}}. \label{eq:0p}
\end{equation} 
Equivalently, $\ket{0}_p$ is a superposition of logical $\ket{+}$ in every displaced GKP basis that has no $p$ modular displacement.

Next, we recognize that CV CPHASE gate is analogous to a qubit CZ gate in the displaced GKP basis,
\begin{eqnarray}\label{eq:czgkp}
   & &\hat{C}_Z^\mathrm{CV} \ket*{l_{\mu_1, \nu_1}^\mathrm{gkp}} \ket*{k_{\mu_2, \nu_2}^\mathrm{gkp}} \nonumber \\
    &=& e^{i\mu_1 \mu_2} (-1)^{lk} \ket*{l_{\mu_1, \nu_1 + \mu_2}^\mathrm{gkp}} \ket*{k_{\mu_2, \nu_2+ \mu_1}^\mathrm{gkp}},
\end{eqnarray}
up to a phase $e^{i\mu_1 \mu_2}$ and a change of modular displacement, both do not alter the quantum information encoded in the displaced GKP basis.  Combining these two facts, an ideal CV cluster state is clearly a superposition of GKP qubit CS with different modular displacement.  This agrees with the observation that qubit CS are ``hidden" in CVCS as derived from a more involved Zak transform \cite{pantaleoni_modular_2020, pantaleoni_hidden_2021, pantaleoni_zak_2023}.  

Finally, we explain the principle behind Steps 2 and 3. To begin, we refer to the one-bit teleportation circuit in Fig.~\ref{fig:combined1}(d), which involves a CNOT gate and a computational basis measurement \cite{10.1103/physreva.62.052316}. In displaced GKP basis, a $q$ displacement by $\sqrt{\pi}$ is a bit-flip up to a phase, i.e. $e^{-i \sqrt{\pi}\hat{p}}\ket*{l^\textrm{gkp}_{\mu,\nu}} = e^{-i\nu \sqrt{\pi}}\ket*{\bar{l}^\textrm{gkp}_{\mu,\nu}}$, where $\bar{l}\equiv l+1 \mod 2$. Therefore, a CD gate implements a CNOT for displaced GKP qubits. The extra phase $e^{-i\nu \sqrt{\pi}}$ is induced only when the ancilla qubit is $\ket{1}$, so it can be corrected by subsequent qubit phase-shift. Furthermore, the $q$-quadrature measurement on the qumode is effectively a computational basis measurement. Explicitly, its outcome, $q=(2m+l)\sqrt{\pi}+\mu$, reveals both the logical value $l$ of the GKP qubit and its modular displacement $\mu$. As such, the circuit of Steps 2 and 3, as shown in Fig.~\ref{fig:combined1}(c), can be understood as a teleportation between the embedded GKP and the ancilla physical qubit.  We note that the post-processing in one-bit teleportation is conditional bit-flip, which is different from the phase-shifts in our protocol. In Appendix we proved that both are equivalent for qubit CS, which the stabilizers equates a bit-flip to a phase-flip of all neighbouring qubits \cite{supp}.

Combining these three facts, the principle of our scheme is transparent: an ideal CVCS is a superposition of qubit CS encoded in the displaced GKP basis, then the qubit-qumode interactions essentially teleports the entanglement in GKP states to the ancilla qubits.

\begin{figure*}
    \centering
    \includegraphics[width=\textwidth]{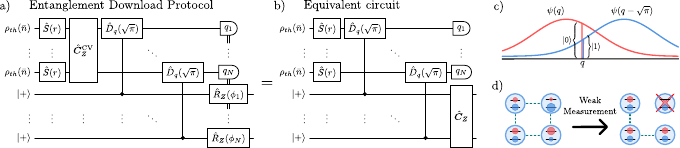}
    \caption{a) Circuit of our protocol when qumodes are prepared as thermal CVCS. 
    b) Equivalent circuit, which each qubit is introduced errors through interacting with squeezed thermal states before being entangled by ideal CZ gates.
    c) Illustration of qubit amplitude imbalance after interacting with finitely squeezed vacuum.
    d) Cluster states after weak measurement correction. Qubits are probabilistically restored to balanced amplitudes or deleted from the cluster.
    }
    \label{fig:combined2}
\end{figure*}

\noindent{\textit{Equivalent circuit model.---}} Preparing ideal CVCS is not physically possible because $\ket{0}_p$ involves infinite energy. Practically, approximate CVCS are realized in experiments that effectively replace the initial $\ket{0}_p$ by finitely squeezed vacuum \cite{menicucci_graphical_2011}. Furthermore, realistic CV platforms must suffer from apparatus artifacts, so the CS generated are usually mixed. Applying our protocol with imperfect CVCS will introduce errors to the downloaded qubit entanglement. Analyzing these errors might be possible by resolving the many-qumode states in the displaced GKP basis, but this could be daunting because the state is mixed and its wavefunction is not flat. Instead, we introduce an elegant equivalent model to map CV imperfections to single-qubit preparation errors.

We first model the imperfect state as a thermal CVCS, where
all qumodes are initialized as identical squeezed thermal states $\rho_{sq} \equiv \hat{S}(r)\rho_{th}(\bar{n})\hat{S}^\dagger(r)$ instead of $\ket{0}_p$. Here $\hat{S}(r) = e^{r( \hat{a}^{\dagger 2}-\hat{a}^{2})/2}$ is the squeezing operator, and $\rho_{th}(\bar{n})$ is a single-mode thermal state with mean excitation $\bar{n}$ \cite{weedbrook_gaussian_2012}. By permuting the CD and CV CPHASE gates, we proved that the qubit state downloaded from an imperfect CVCS is equivalent to a qubit CS with the same graph, except that every qubit is not initialized as $\ket{+}$ but error is introduced through a CD gate with squeezed thermal state followed by $q$-measurement \cite{supp}.
The equivalence is illustrated in Figs.~\ref{fig:combined2}(a) and (b), or mathematically as
\begin{eqnarray}
    && \RZ \bra{\bm{q}} \hat{\bm{C}}_D \CPHASE ( \rho_{sq} \otimes \ketbra{+}{+} )^{\otimes N}\CPHASEdag \CD^\dagger \ket{\bm{q}} \RZ^\dagger \nonumber \\
    &=\, \, & \CZ \bigotimes_{i = 1}^N \bra{q_i}_q \hat{C}_{D} (\rho_{sq} \otimes \ketbra{+}{+}) \hat{C}_{D}^\dagger \ket{q_i}_q \CZdag . \label{eqn:ECM}
\end{eqnarray}

To analyze the qubit errors introduced by CV imperfections, we first recall that a squeezed thermal state can be treated as a mixture of $p$-displaced squeezed states, i.e. $\rho_{sq} = \int_{-\infty}^{\infty} dp_0 \, f(p_0, \sigma^2) \hat{D}_p(p_0) \hat{S}(r_0) \ketbra{\mathrm{vac}}{\mathrm{vac}} \hat{S}^\dagger(r_0) \hat{D}_p^\dagger(p_0)$, where $f(p_0, \sigma^2) \equiv e^{-p_0^2/(2\sigma^2)}/\sqrt{2\pi \sigma^2}$ is a Gaussian distribution of displacement $p_0$ with variance $\sigma^2 = (\bar{n} + \bar{n}^2)/(e^{2r}(1 + 2\bar{n}))$ \cite{leonhardt_essential_2010, supp}, and the effective squeezing is $e^{2r_0} = e^{2r}(1 + 2\bar{n})$. For each $p$-displaced squeezed state, the qubit state becomes
\begin{eqnarray}
\ket{\Psi(q)} &\propto& \bra{q}_q \hat{C}_D \hat{D}(p_0) \hat{S}(r_0)\ket{\mathrm{vac}} \ket{+} \nonumber \\
&\propto&   \psi(q)\ket{0} + e^{-ip_0\sqrt{\pi}}\psi(q-\sqrt{\pi})\ket{1}, \label{eqn:qubitcd}    
\end{eqnarray}
where $\psi(q) \equiv \bra{q} \hat{S}(r_0) \ket{\mathrm{vac}}$ is the wavefunction of finitely squeezed vacuum state. Eq.~(\ref{eqn:qubitcd}) reveals that CV imperfections introduce two types of qubit errors. First, since $\psi(q)$ is not flat due to finite squeezing, as illustrated in Fig.~\ref{fig:combined2}(c), the qubit becomes an unequal superposition of $\ket{0}$ and $\ket{1}$ as opposed to an equal superposition in ideal qubit CS. Second, the $p$-displacement induces a phase shift on the qubit.  Since the displacement is a random variable, it leads to qubit dephasing.

\noindent{\textit{Finite squeezing and probabilistic correction.---}} Momentarily, we focus on the finite squeezing error and neglect thermalization, i.e. assume $\bar{n}=0$ and thus $p_0=0$. We first note that the amplitude imbalance $\gamma \equiv |\psi(q-\sqrt{\pi})/\psi(q)|$ is known because both the measurement outcome $q$ and initial squeezing $r$ are known. Correcting a known amplitude imbalance is possible probabilistically by weakly measuring the qubit with positive operator-valued measures (POVM) $\{\hat{M}_0, \hat{M}_1\} \equiv \{\gamma \ketbra{0}{0} + \ketbra{1}{1}, \sqrt{1 - \gamma^2}\ketbra{0}{0}\}$ if $\gamma < 1$, and  $\{\hat{M}_0, \hat{M}_1\}\equiv \{ \ketbra{0}{0} + \gamma^{-1} \ketbra{1}{1}, \sqrt{1 - \gamma^{-2}}\ketbra{1}{1}\}$ if $\gamma > 1$.  As illustrated in Fig.~\ref{fig:combined2}(d), if $\hat{M}_0$ is obtained, the qubit is restored to equal superposition; however, if $\hat{M}_1$ is obtained, the qubit is projected to a computational basis state, which deletes the qubit from the CS \cite{raussendorf_one-way_2001}. Since the POVM commute with the qubit CZ gates in the equivalent circuit Eq.~(\ref{eqn:ECM}), the weak measurement can be conducted after entanglement is downloaded to qubits.

Averaging over the probability of $q$-measurement outcome, the average probability of failed correction, hence qubit deletion, is
\begin{equation}\label{eq:pdel}
    p_\textrm{del} = \text{erf}(e^{-r_0}\sqrt{\pi}/2)~.
\end{equation}
Our correction scheme turns the finite squeezing error to qubit erasure error. This error has attracted much attention recently because high threshold quantum error correction (QEC) codes were found \cite{kubica_erasure_2023, grassl_codes_1997, knill_scalable_2005, wu_erasure_2022, levine_demonstrating_2024}. Particularly, for FTQC, schemes based on surface code can tolerate up to $24.9\%$ of qubit loss \cite{barrett_fault_2010}. This is equivalent to $11.9$ dB squeezing according to Eq.~(\ref{eq:pdel}).
We note that this squeezing threshold cannot be directly compared with the $\approx 10$~dB squeezing threshold of CV-based QC in the literature \cite{menicucci_fault-tolerant_2014, fukui_high-threshold_2018,bourassa_blueprint_2021,tzitrin_fault-tolerant_2021, larsen_fault-tolerant_2021}. While the literature usually considers the squeezing threshold of highly non-Gaussian GKP ancilla, our threshold is that of the Gaussian CVCS, achieving which can be much more practical.

Apart from FTQC, CS can also be used for simple (non-FT) QC or as quantum memory. For these applications, respective schemes have been developed that can tolerate up to $50\%$ qubit loss \cite{varnava_loss_2006, stace_thresholds_2009}, which corresponds to $5.4$~dB squeezing. To the best of our knowledge, these are the first squeezing thresholds found for applications other than FTQC. This level of squeezing is close to that in recent experiments \cite{asavanant_generation_2019, larsen_deterministic_2019}, if not already realized \cite{10.1063/5.0144385}.
The squeezing threshold could be further suppressed by using CS in which each vertex is represented by interconnected qumodes, analogous to repeater CS \cite{azuma_all-photonic_2015,supp}.

\begin{figure*}
    \centering    \includegraphics[width=\textwidth]{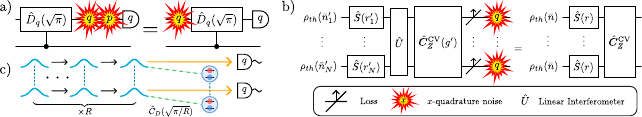}
    \caption{a) Detector inefficiency introduces both $q$ and $p$ quadrature noises. $p$-noise is irrelevant after $q$-measurement, while $q$-noise can be lumped into CVCS imperfection.
    b) Correlation induced by apparatus artifacts can be canceled by introducing mode-dependent thermalization and squeezing, additional beam-splitter array, and modifying strength of CPHASE, i.e. $\CPHASE(g') \equiv \prod_{(i,j) \in E} e^{ig'\hat{q}_i \hat{q}_j}$.
    c) Weak CD strength can be compensated by interacting the qubits with multiple copies of CVCS.
    }
    \label{fig:combined3}
\end{figure*}

\noindent{\textit{Thermalization, loss, and detection inefficiency.---}} Now we return to the apparatus artifacts. In our model, we already considered the initial thermalization, which appears when, e.g. the qumodes are initially not sufficiently cooled. 
To verify our previous hint that thermalization causes dephasing on the downloaded qubit state, we average the Gaussian distributed displacement $p_0$ in Eq.~(\ref{eqn:qubitcd}). The transformation of each qubit is found as
\begin{equation}
    \ket{+} \xrightarrow{} (1-p_\Phi)\ketbra{\Psi(q)} + p_\Phi\hat{Z} \ketbra{\Psi(q)} \hat{Z}, 
\end{equation}
which corresponds to a dephasing channel with dephasing rate $p_\Phi = (1 - e^{-\pi \sigma^2/2})/2$.  We note that since the amplitude-balancing POVM are diagonal in computational basis, our weak measurement correction scheme is still functional under dephasing.

Apart from thermalization, channel loss and detector inefficiency are also major artifacts in CV platforms. For channel loss, we specifically consider the loss when CVCS is transported to interact with the qubits. 
For detector inefficiency, we assume that it is well characterized, so compensating amplification is applied. The total effect is equivalent to the qumode experiencing random displacement in both $q$ and $p$ quadratures before ideal detection \cite{2018PhRvA..97c2346A}. However, not both of the noises aﬀect the downloaded entanglement. Specifically, $p$-displacement noise is represented by a displacement operation $e^{i \varepsilon \hat{q}}$ with unknown $\varepsilon$, so it becomes an unimportant global phase after a projective measurement in $q$ quadrature. Whereas the $q$-displacement noise does reduce entanglement; it commutes with CD gate and can be lumped as CVCS imperfection (see Fig.~\ref{fig:combined3}a). Overall, the errors introduced by channel loss and detector inefficiency can be treated as qubits downloading entanglement from an imperfect CVCS.

Nevertheless, these errors are generally not equivalent to thermalization. Intuitively, since CPHASE gates establish correlations between qumodes, noise introduced to one qumode in CVCS should be equivalent to correlated noise in multiple qumodes before being entangled by CPHASE.
According to equivalent circuit model, this implies that these errors do not become dephasing on individual qubits, instead they cause correlated dephasing among multiple qubits. Correlated noise is usually difficult to handle by the QEC schemes. 

To tackle unwanted correlation, a simple resolution is to counteract with additional (anti-)correlation. Correlation can be introduced efficiently in CV systems as it requires only Gaussian operations. Specifically, our aim is to construct a many-qumode entangled state that becomes a thermal CVCS after suffering from channel loss and detector noise, so the downloaded error remains individiual qubit dephasing. As shown in Fig.~\ref{fig:combined3}(b), one possible construction is introducing mode-dependent squeezing and thermalization, an additional array of beam-splitters, and CPHASE gates with modified strength $g'$. At the leading order of loss rate $\epsilon_1$ and detector inefficiency $\epsilon_2$, we can construct a thermal CVCS with
\begin{eqnarray}
    e^{2r} &\approx & e^{2r'} - \epsilon_1 \mathcal{W}_- -\epsilon_2 \mathcal{V}_-, \nonumber\\
    \bar{n} &\approx & \frac{\epsilon_1}{2}\qty( \mathcal{W}_+ e^{-2r'} - 1)+ \frac{\epsilon_2}{2}\qty(\mathcal{V}_+ e^{-2r'}) ,\\
    g' &\approx & 1 + e^{-2r'} (\epsilon_1 + 2\epsilon_2)\nonumber
\end{eqnarray}
where $\mathcal{W}_{\pm} \equiv (d^2 e^{4r'} + e^{4r'} \pm 1)/2$ and $\mathcal{V}_{\pm} \equiv d^2 e^{4r'} \pm 1$; $d$ is the maximum degree of graph $G$ \cite{biggs_algebraic_1974}; $r'$ is the maximum achievable squeezing of the initial qumodes.

\noindent{\textit{Weak interaction.---}} Step 2 of our protocol assumed the qumode is displaced by $\sqrt{\pi}$ conditionally on the qubit state, but realizing an interaction with such strength may be challenging in some platforms. Fortunately, this requirement is not necessary. If each CD gate can only produce a weak conditional displacement of amplitude $\sqrt{\pi/R}$, then a qubit CS can still be downloaded by interacting with $R$ copies of CVCS, as illustrated in Fig.~\ref{fig:combined3}(c).

One may worry that this approach may worsen the imperfection noise; interestingly, this is not true. Consider the total CD operation applied on a qubit, $\hat{I} \otimes \ketbra{0}{0} + \exp(-i \sqrt{\frac{\pi}{R}} \sum_{s=1}^R  \hat{p}^{(s)}) \otimes \ketbra{1}{1}$, where the superscript denotes the $s$th qumode copy. The operation can be seen as a CD gate with strength $\sqrt{\pi}$ applied between the qubit and the collective mode $\hat{A}= \sum_{s=1}^R  \hat{a}^{(s)}/\sqrt{R}$. It is easy to check from its covariance matrix that the state of the collective modes is identical to each copy of CVCS \cite{supp}. Therefore, the downloaded error of this protocol is identical to the original version, albeit using weaker CD gates.

\noindent{\textit{Physical Implementation.---}} Our protocol is implementable with four ingredients that are available on many hybrid qubit-CV quantum platforms: CV cluster states, ancilla qubits, CD gate, and $q$-quadrature measurement. Here, we suggest some explicit examples.

In quantum optical platforms, CVCS can be efficiently generated among temporal or frequency modes \cite{yokoyama_ultra-large-scale_2013, chen_experimental_2014}, and $q$-quadrature measurement can be implemented by homodyne detection \cite{weedbrook_gaussian_2012}. One possible choice of qubit is an atom trapped in an optical cavity. The CD gates can then be realized by combining atomic state-dependent phase rotation of the optical state with strong linear displacement \cite{wang_engineering_2005,hastrup_protocol_2022,dhara_interfacing_2024}. Another possibility of qubit would be traveling electrons prepared in a frequency comb \cite{dahan_creation_2023, baranes_free-electron_2023}. The scattering between the electronic comb state and photonic modes resembles a CD gate.

Another possible implementation would be superconducting circuit quantum electrodynamics (cQED) systems. CVCS have been realized with frequency modes in a resonator \cite{jolin_multipartite_2023, petrovnin_generation_2023, hernandez_control_2024}, and qubits can be the transmon qubits coupled to it. CD gate can be realized by combining the intrinsic dispersive coupling with linear drive \cite{eickbusch_fast_2022}. Although homodyne detection is not intrinsic in cQED platforms, $q$-quadrature measurement can be performed by repeated measurement of a coupling auxiliary qubit \cite{strandberg_digital_2024} or phase estimation \cite{terhal_encoding_2016}.

\noindent{\textit{Conclusion.--}} We present a novel top-down approach to generate many-qubit entanglement: by downloading it from the efficiently generated CV cluster states. Our scheme can be implemented with operations available in many hybrid quantum systems. We introduce an equivalent circuit model to map dominant CV errors to single-qubit preparation errors, so they can be mitigated through established qubit QEC. We expect our result can facilitate the generation of resourceful entanglement for various quantum technologies, and introduce new applications of hybrid qubit-CV platforms \cite{Andersen:2015dp,liu_hybrid_2024}.

This work is supported by the Natural Sciences and Engineering Research Council of Canada (NSERC) Discovery Grant (RGPIN-2021-02637) and CREATE (543245-2020-CREAT) and Canada Research Chairs (CRC-2020-00134). 


\let\oldaddcontentsline\addcontentsline
\renewcommand{\addcontentsline}[3]{}
\bibliographystyle{apsrev4-2}
\bibliography{references}

\let\addcontentsline\oldaddcontentsline

\clearpage

\onecolumngrid
\newpage
\begin{center}
\vspace{2em}
\textbf{\large Supplemental Information for:  Downloading many-body entanglement from a continuous variable cluster state to a qubit cluster state}

\vspace{1em}

\text{Zhihua Han\textsuperscript{1} and Hoi-Kwan Lau\textsuperscript{1}}

\vspace{0.2em}

\textit{\small \textsuperscript{1}\SFU}\\
\text{\small (Dated: \today)}

\end{center}

\setcounter{equation}{0}
\setcounter{figure}{0}
\setcounter{table}{0}
\setcounter{section}{0}
\setcounter{page}{1}
\renewcommand{\thefigure}{S\arabic{figure}}
\renewcommand{\theHfigure}{S\arabic{figure}}
\renewcommand{\theequation}{S\arabic{equation}}
\onecolumngrid


\tableofcontents

\section{Proof of Equations 1 and 4} \label{chap:ECM}

\subsection{Proof of Equation 4}

We first show the general case of Eq. (4), which will yield Eq. (1) as a special case. Conditional displacement on all modes is defined as 
\begin{equation}
\CD= \prod_{i = 1}^N \hat{C}_{D_{i}}, \quad 
\hat{C}_D \equiv \hat{I} \otimes \ketbra{0}{0} + e^{-i\sqrt{\pi}\hat{p}}\otimes \ketbra{1}{1} = \exp(-i\sqrt{\pi}\hat{p} \hat{b}),
\end{equation}
where the bit-operator $\hat{b}$ on qubit $i$ is given by
\begin{equation}
    \hat{b}_i \equiv \frac{1 - \hat{Z}_i}{2}, \label{eqn:bit_2}
\end{equation}
which $\hat{b}_i \ket{0} = 0$ and $\hat{b}_i \ket{1} = \ket{1}$.

To prove Eq. (4) in the main text, we first commute the CPHASE gates with conditional displacement by using the identity
\begin{align}
    \begin{split}
    \CD \CPHASE \CD^\dagger &= \exp(\frac{i}{2}\CD \Q^T \mathbf{A} \Q  \CD^\dagger) = \exp(\frac{i}{2} (\Q - \hat{\bm{b}} \sqrt{\pi})^T \mathbf{A} (\Q - \hat{\bm{b}} \sqrt{\pi}) ), \\ 
    &= \underbrace{\exp(\frac{i}{2} \Q^T \mathbf{A} \Q)}_{\CPHASE}\underbrace{\exp(-i \sqrt{\pi} \hat{\bm{b}}^T \mathbf{A} \Q )}_{\text{mode-qubit coupling}} \underbrace{\exp(\frac{i\pi}{2} \hat{\bm{b}}^T \mathbf{A} \hat{\bm{b}})}_{\CZ},
    \end{split} 
    \label{eqn:ECMpart2}
\end{align}
where we have used the relation $\hat{C}_D\hat{q} \hat{C}_D^\dagger = \hat{q} - \hat{b} \sqrt{\pi}$, and defined $\hat{\bm{b}} \equiv \mqty(\hat b_1 & \ldots &\hat{b}_N)$, $\hat{\bm{q}} \equiv \mqty(\hat q_1 & \ldots &\hat{q}_N)$.

In the second line of \cref{eqn:ECMpart2}, we decompose the operators into three collections of like operators: CPHASE gates, qubit CZ gates, and mode-qubit coupling that will later reduce to relative phase shifts. Combining with the projector that represents the $q$ quadrature measurement of all modes, we have
\begin{align}
    \bra{\bm{q}}_q \CD \CPHASE &=  \bra{\bm{q}}_q  \CPHASE e^{-i \sqrt{\pi} \hat{\bm{b}}^T \mathbf{A} \Q} \CZ \CD.  \label{eqn:ECMderiv-3}
\end{align}

These operators can be simplified upon the projector. First, the CPHASE gates $\CPHASE$ will become a global phase:
\begin{equation}
    \bra{\bm{q}}_q \CPHASE  = \exp(\frac{i}{2}\bm{q}^T\mathbf{A}\bm{q}) \bra{\bm{q}}_q. \label{eqn:ECMderiv.1}
\end{equation}
Second, the mode-qubit coupling $e^{-i \sqrt{\pi} \hat{\bm{b}}^T \mathbf{A} \Q}$ will become phase shifts on the qubits, $\RZ(-\bm{\phi})$, which the phase angles are known from the $q$ measurement outcomes: 
\begin{equation}
    \bra{\bm{q}}_q e^{-i \sqrt{\pi} \hat{\bm{b}}^T \mathbf{A} \hat{\bm{q}} } =  e^{-i \sqrt{\pi} \hat{\bm{b}}^T \mathbf{A} \bm{q} } \bra{\bm{q}}_q = \RZ(-\bm{\phi}) \bra{\bm{q}}_q. \label{eqn:ECMderiv.2}
\end{equation}
The phase shift correction $\bm{\phi}$ is given by
\begin{equation}
    \phi_i = \sqrt{\pi} (\mathbf{A }\bm{q})_i = \sqrt{\pi} \sum_{j \in n(i)}  q_j, \label{eqn:phidef}
\end{equation}
which depends on the measured $q$ quadratures of all modes adjacent to mode $i$. 

The third operator is the qubit CZ gates, which acts only on the ancilla degree of freedom and thus commutes with $q$ quadrature measurement:
\begin{equation}
    \bra{\bm{q}}_q \CZ  = \CZ \bra{\bm{q}}_q. \label{eqn:ECMderiv.3}
\end{equation}

Combining all of these results, \cref{eqn:ECMderiv-3} becomes
\begin{align}
    \bra{\bm{q}}_q \CD \CPHASE &= \RZ(-\bm{\phi}) \CZ \bra{\bm{q}}_q \CD.  \label{eqn:ECMabstract}
\end{align}
When the initial qumode-qubit state is $(\rho_{sq} \otimes \ketbra{+}{+})^{\otimes N}$ in \cref{eqn:ECMabstract}, after applying the phase correction $\RZ(\bm{\phi})$ on both sides, we obtain Eq. (4) of the main text:
\begin{equation}
    \RZ(\bm{\phi}) \bra{\bm{q}}_q \hat{\bm{C}}_D \CPHASE ( \rho_{sq} \otimes \ketbra{+}{+} )^{\otimes N}\CPHASEdag \CD^\dagger \ket{\bm{q}}_q \RZ^\dagger(\bm{\phi}) 
    = \CZ \bigotimes_{i = 1}^N \bra{q_i}_q \hat{C}_{D} (\rho_{sq} \otimes \ketbra{+}{+}) \hat{C}_{D}^\dagger \ket{q_i}_q \CZdag.
    \label{eqn:ECMthermal}
\end{equation}

\subsection{Proof of Equation 1}

We obtain the special case of Eq. (1) when the initial state is an ideal CV cluster state, i.e. when the initial mode is a zero $p$ eigenstate, $\rho_{sq} = \ketbra{0}{0}_p$. We define $\ket*{G^\mathrm{CV}} \equiv \CZ^{\mathrm{CV}} \ket{0}_p^{\otimes N}$, $\ket{G} \equiv \CZ \ket{+}^{\otimes N}$, so \cref{eqn:ECMthermal} becomes
\begin{equation}
    \RZ(\bm{\phi}) \bra{\bm{q}}_q \hat{\bm{C}}_D  \qty(\ketbra*{G^{\mathrm{CV}}}{G^{\mathrm{CV}}} \otimes \ketbra{+}{+}^{\otimes N} ) \CD^\dagger \ket{\bm{q}}_q \RZ^\dagger(\bm{\phi})=\CZ \bigotimes_{i = 1}^N \bra{q_i}_q \hat{C}_{D} (\ketbra{0}{0}_p \otimes \ketbra{+}{+}) \hat{C}_{D}^\dagger \ket{q_i}_q \CZdag. \label{eqn:ECM_Gsub}
\end{equation}
Then, for a single mode to single qubit, we have the following identity:
\begin{equation}
    \bra{q}_q \hat{C}_D \ket{0}_p \ket{+} = \frac{1}{\sqrt{2}}\bra{q}_q \qty(\ket{0}_p \ket{0} + e^{-i\sqrt{\pi}\hat{p}}\ket{0}_p \ket{1}) \propto \ket{+}, \label{eqn:ECMsinglemodeidentity}
\end{equation}
since $e^{-i\sqrt{\pi}\hat{p}}\ket{0}_p = \ket{0}_p$. Substituting \cref{eqn:ECMsinglemodeidentity} into \cref{eqn:ECM_Gsub}, we obtain 
\begin{equation}
    \RZ(\bm{\phi}) \bra{\bm{q}}_q \hat{\bm{C}}_D  \qty(\ketbra*{G^{\mathrm{CV}}}{G^{\mathrm{CV}}} \otimes \ketbra{+}{+}^{\otimes N} ) \CD^\dagger \ket{\bm{q}}_q \RZ^\dagger(\bm{\phi}) \propto \ketbra{G}{G}.
\end{equation}
This is the density matrix form of Eq. (1) of the main text:
\begin{equation}
    \RZ(\bm{\phi}) \bra{\bm{q}}_q\CD \ket*{G^\mathrm{CV}} \ket{+}^{\otimes N} \propto \ket{G}~.
\end{equation}

\section{Proof of Equation 2}\label{chap:gkp}

In this section, we show that an ideal zero-momentum eigenstate can be expressed as a superposition of displaced GKP plus states $\ket*{+_{\mu, 0}^\mathrm{gkp}}$. Consider

\begin{align}
\begin{split}
        \ket{0}_p \propto \int_{-\infty}^\infty \ket{q}_q dq & = \sum_{m \in \mathbb{Z}} \int_{-\sqrt{\pi}/2}^{\sqrt{\pi}/2} d\mu \ket{m\sqrt{\pi}+\mu}_q,  \\
    &= \sum_{m \, \text{even}}  \int_{-\sqrt{\pi}/2}^{\sqrt{\pi}/2} d\mu \ket{m\sqrt{\pi}+\mu}_q  + \sum_{m \, \text{odd}}  \int_{-\sqrt{\pi}/2}^{\sqrt{\pi}/2} d\mu \ket{m\sqrt{\pi}+\mu}_q, \\
    &\propto \int_{-\sqrt{\pi}/2}^{\sqrt{\pi}/2} d\mu \, \ket*{0^\mathrm{gkp}_{\mu, 0}} + \ket*{1^\mathrm{gkp}_{\mu, 0}}, \label{paper:eqn2}
\end{split}
\end{align}
where $\ket*{0^\mathrm{gkp}_{\mu, 0}} =  \sum_{m\,  \text{even}} \hat{D}_q(\mu) \ket{m\sqrt{\pi} }_q = \sum_{m\,  \text{even}} \ket{m\sqrt{\pi} + \mu}_q$, $\ket*{1^\mathrm{gkp}_{\mu, 0}} =  \sum_{m\,  \text{odd}} \hat{D}_q(\mu) \ket{m\sqrt{\pi}}_q = \sum_{m\,  \text{odd}} \ket{m\sqrt{\pi} + \mu}_q$. This yields Eq. (2) in the main text.
 

\section{Proof of Equation 3} 
\label{sec:CPHASEdGKP}

Eq. (3) in the main text can be computed directly:

\begin{align}
    \begin{split}
    &e^{i\hat q_1 \hat q_2} \ket*{l^\mathrm{gkp}_{\mu_1, \nu_1}} \ket*{k^\mathrm{gkp}_{\mu_2, \nu_2}},\\ &= \sum_{m_1, m_2} e^{i((2m_1 + l)\sqrt{\pi}+ \mu_1)((2m_2 + k)\sqrt{\pi}+ \mu_2)} e^{i\nu_1(2m_1 + l)\sqrt{\pi}} e^{i\nu_2(2m_2 + k)\sqrt{\pi}} \ket{(2m_1 + l)\sqrt{\pi} + \mu_1}_q \ket{(2m_2 + k)\sqrt{\pi} + \mu_2}_q, \\
    &= e^{i\mu_1 \mu_2}(-1)^{lk} \sum_{m_1, m_2} e^{i(\nu_1 + \mu_2)(2m_1 + l)\sqrt{\pi}} e^{i(\nu_2 + \mu_1)(2m_2 + k)\sqrt{\pi}} \ket{(2m_1 + l)\sqrt{\pi} + \mu_1}_q \ket{(2m_2 + k)\sqrt{\pi} + \mu_2}_q, \\
    &= e^{i\mu_1 \mu_2} (-1)^{lk} \ket*{l_{\mu_1, \nu_1 + \mu_2}^\mathrm{gkp}} \ket*{k_{\mu_2, \nu_2+ \mu_1}^\mathrm{gkp}}.
    \end{split} 
    \label{eqn:CZgkpdisp_twomode}
\end{align}
where we have used the fact that $\exp(i m \pi) = (-1)^m$ for any integer $m$.

\subsection{Multi-mode generalization}

By repeatedly applying \cref{eqn:CZgkpdisp_twomode}, it is easy to show that applying a product of CPHASE gates to an $N$-mode displaced GKP state with displacements $(\bm{\mu}, \bm{\nu})$ will generate an $N$-mode GKP cluster state with the new displacements $(\bm{\mu}, \bm{\nu} + \mathbf{A} \bm{\mu})$, i.e.
\begin{equation}
    \CPHASE(\mathbf{A}) \ket*{\bm{+}_{\bm{\mu}, \bm{\nu} }^\mathrm{gkp}}= e^{i \bm{\mu}^T \mathbf{A}\bm{\mu}} \CZ^\mathrm{gkp}(\mathbf{A}) \ket*{\bm{+}^\mathrm{gkp}_{\bm{\mu}, \bm{\nu} + \mathbf{A} \bm{\mu}}}, \label{eqn:GKPCZgate}
\end{equation}
where $e^{i \bm{\mu}^T \mathbf{A}\bm{\mu}}$ is a global phase, and we have defined the $N$-mode displaced GKP plus state as $| \bm{+}^\mathrm{gkp}_{\bm{\mu}, \bm{\nu}}\rangle \equiv  |+^\mathrm{gkp}_{\mu_1,\nu_1}\rangle \otimes \ldots |+^\mathrm{gkp}_{\mu_N,\nu_N}\rangle$. The product of GKP CZ gates in \cref{eqn:GKPCZgate} is defined as 
\begin{equation}
    \CZ^\mathrm{gkp}(\mathbf{A}) \equiv \prod_{(i, j) \in E} \hat{C}_{Z_{ij}}^\mathrm{gkp}, \label{eqn:gkpCZadj}
\end{equation}
and each GKP CZ gate is defined as
\begin{equation}
\hat{C}_Z^\mathrm{gkp} \ket*{{l_1}_{\mu_1, \nu_1}^\mathrm{gkp}} \ket*{{l_2}_{\mu_2, \nu_2}^\mathrm{gkp}} \equiv (-1)^{l_1 l_2}\ket*{{l_1}_{\mu_1, \nu_1}^\mathrm{gkp}} \ket*{{l_2}_{\mu_2, \nu_2}^\mathrm{gkp}}, \label{eqn:gkpCZ}
\end{equation}
which applies the conditional phase flip but does not change the displacement basis. 

Combining \cref{paper:eqn2} and \cref{eqn:GKPCZgate}, we find that an ideal CV cluster state is a superposition of qubit cluster states in the displaced GKP basis,
\begin{align}
    \ket*{G^\mathrm{CV}} 
    = \CPHASE(\mathbf{A}) \ket{0}_p^{\otimes N} = \frac{1}{\pi^{N/2}} \int_{-\sqrt{\pi}/2}^{\sqrt{\pi}/2} d \bm{\mu} \, e^{i \bm{\mu}^T \mathbf{A}\bm{\mu}} |G^\mathrm{gkp}_{\bm{\mu},\bm{A\mu}}\rangle ,\label{eqn:dispGKPcluster}
\end{align}
where we let $\int d\bm{\mu} \equiv \int \ldots \int d\mu_1\ldots d\mu_{q_N}$, and all integral domains are $\mu_i \in [-\sqrt{\pi}/2, \sqrt{\pi}/2)$. 
\section{Hybrid one-bit teleportation}
\subsection{Conditional displacement} \label{sec:XZdispGKP}

When we apply displacement operators to a displaced GKP state, we find that its logical value is flipped while a phase is generally induced:
\begin{align}
\begin{split}
    \hat{D}_q(\sqrt{\pi}) \ket*{l_{\mu, \nu}^\mathrm{gkp}} = e^{-i\nu \sqrt{\pi}} \ket*{\bar{l}_{\mu, \nu}^\mathrm{gkp}},
\end{split}\label{eqn:CXdispgkp}
\end{align}
where $l \in \{0, 1\}$ and $\bar{l} \equiv l + 1 \pmod 2$. This implies that the logical X gate of the displaced GKP basis is 
\begin{equation}
    \hat{X}^\mathrm{gkp}_{\mu, \nu} \equiv e^{i\nu \sqrt{\pi}} \Dq{\sqrt{\pi}} = e^{i\nu \sqrt{\pi}} e^{-i \sqrt{\pi} \hat{p}} \label{eqn:displacedGKPgates}.
\end{equation}
Therefore, when we apply conditional displacement gate to a displaced GKP state, we find  
\begin{align}
    \begin{split}
        \hat{C}_D(\sqrt{\pi})|l_{\mu, \nu}^{\mathrm{gkp}}\rangle\ket{0} &= |l_{\mu, \nu}^{\mathrm{gkp}}\rangle  \ket{0}, \\
        \hat{C}_D(\sqrt{\pi})|l_{\mu, \nu}^{\mathrm{gkp}}\rangle\ket{1} &=
    e^{-i\nu\sqrt{\pi}}|\bar{l}_{\mu, \nu}^{\mathrm{gkp}} \rangle  \ket{1}. 
    \end{split}
    \label{eqn:CDdispGKP}
\end{align}
This is equivalent to a logical CNOT gate on the displaced GKP state with an additional phase shift $\hat{R}_Z(-\nu\sqrt{\pi})$ on the auxiliary qubit. 
 
Now consider an arbitrary qubit state encoded in a particular displaced GKP basis $\ket*{\psi_{\mu, \nu}^\mathrm{gkp}} \equiv \alpha \ket*{0_{\mu, \nu}^\mathrm{gkp}} + \beta \ket*{1_{\mu, \nu}^\mathrm{gkp}}$. We now implement one-bit teleportation \cite{10.1103/physreva.62.052316, nielsen_cluster-state_2006} in \cref{fig:onebitgeneralizedZ} by using a conditional displacement as a logical CNOT gate, as illustrated in \cref{fig:onebitequal}. Because a conditional displacement implements both a logical CNOT gate and an ancilla qubit rotation gate $\hat{R}_Z(\nu \sqrt{\pi})$, a counteractive ancilla phase shift has to be applied to implement the teleportation.

\begin{figure}[htpb]
    \centering
        \begin{quantikz}[row sep={0.9cm,between origins}] 
    \lstick{$\ket*{\psi}$} &\targ{} & \meter{l} \\
    \lstick{$\ket{+}$} &  \ctrl{-1}  \qw   &  \gate{\hat{X}^l} \wire[u][1]{c} & \rstick{$\ket*{\psi}$} 
\end{quantikz}=\begin{quantikz}[row sep={0.9cm,between origins}] 
    \lstick{$\ket*{\psi}$} &\targ{} & &\gategroup[2,steps=2,style={dashed,rounded
        corners, inner
xsep=2pt,color=red},background,label style={label
position=below,anchor=mid,yshift=-0.6cm}]{post-processing} & \meter{} \\
    \lstick{$\ket{+}$} &  \ctrl{-1}  \qw   &  \gate{\hat{R}_Z(-\theta)} & \gate{\hat{R}_Z(\theta)} & \gate{\hat{X}^l } \wire[u][1]{c}& \rstick{$  \ket*{\psi}$} 
\end{quantikz}
    \caption[One-bit teleportation circuit.]{Left: One-bit teleportation circuit. Right: One-bit teleportation circuit with extra $\hat{R}_Z$ gates added. $l \in \{0, 1\}$ is the logical value measured from the qubit. We have defined $\ket*{\psi} \equiv \alpha \ket*{0} + \beta \ket*{1}$ and added the identity gate as a combination of $\hat{R}_Z$ gates for an easier comparison with the one-bit teleportation in displaced GKP states.}
    \label{fig:onebitgeneralizedZ}
\end{figure}

\begin{figure}[htpb]
    \centering
    \begin{quantikz}[row sep={1cm,between origins}]
    \lstick{$\ket*{\psi^\mathrm{gkp}_{\mu, \nu}}$} & \targ{} \gategroup[2,steps=2,style={dashed,rounded
        corners, inner
xsep=2pt,color=ao},background,label style={label
position=below,anchor=mid,yshift=-0.6cm}]{$\hat{C}_D$ in disp. GKP basis }  & \qw &\gategroup[2,steps=2,style={dashed,rounded
        corners, inner
xsep=2pt,color=red},background,label style={label
position=below,anchor=mid,yshift=-0.6cm}]{post-processing} &  \meterD{\mu, l}  \\
    \lstick{$\ket{+}$} &\ctrl{-1} &  \gate{\hat{R}_Z(-\nu\sqrt{\pi})} & \gate{\hat{R}_Z(\nu\sqrt{\pi})}  &\gate{\hat{X}^l} \wire[u][1]{c}  &  \rstick{$  \ket*{\psi}$} 
\end{quantikz}=\begin{quantikz}[row sep={1cm,between origins}]
       \lstick{$\ket*{\psi_{\mu, \nu}^\mathrm{gkp}}$} & \gate{\hat{D}_q(\sqrt{\pi})} \gategroup[2,steps=1,style={dashed,rounded
        corners, inner
xsep=2pt,color=ao},background,label style={label
position=below,anchor=mid,yshift=-0.6cm}]{$\hat{C}_D$}& \meterD{q} \gategroup[2,steps=1,style={dashed,rounded
        corners, inner
xsep=46pt,color=red,xshift=1.6cm},background,label style={label
position=below,anchor=mid,yshift=-0.6cm}]{by-product}  \\
       \lstick{$\ket{+}$} &\ctrl{-1} &    \rstick{$\hat{R}_Z(-\nu\sqrt{\pi})\hat{X}^l\ket{\psi}$} 
\end{quantikz}
    \caption[Hybrid one-bit teleportation on a displaced GKP state.]{Left: Conditional displacement is a CNOT gate in the displaced GKP basis plus a phase shift on the physical qubit. Since $q = (2m+l)\sqrt{\pi} + \mu$, we can express the qumode measurement outcome and post-processing in terms of $l, \nu$. Right: One-bit teleportation without post-processing, so the teleported state will contain the teleportation by-product.}
\label{fig:onebitequal}
\end{figure}

To do so, we need to know the modular $p$ displacement, $\nu$, of the displaced GKP states. As shown in \cref{eqn:dispGKPcluster}, an ideal CV cluster state is a superposition of displaced GKP cluster states with modular displacements $(\bm{\mu}, \mathbf{A}\bm{\mu})$. Measuring all modes in $q$ quadrature will collapse the displacement superposition and reveal the value of $\bm{\mu}$, and hence the modular $p$ displacement, $\bm{\nu}=\mathbf{A}\bm{\mu}$. By using the explicit form of adjacency matrix $\mathbf{A}$, the phase shift $\theta_i$ that needs to be corrected on qubit $i$ is given by
\begin{equation}
    \theta_i = \sqrt{\pi} (\mathbf{A \bm{\mu}})_i = \sqrt{\pi} \sum_{j \in n(i)}  \mu_j,\label{eqn:thetadef}
\end{equation}
which depends on the $\mu$'s of all adjacent modes. As shown in \cref{fig:onebitequal}, the overall post-processing requires a phase-shift that depends on the measured modular displacement and a bit flip that depends on the measured GKP qubit logical value. Initially, this post-processing seems different than the post-processing in our scheme as in \cref{eqn:phidef}, which is a phase shift that depends on the measured quadrature value. However, we will show in the following section that they are the same for an ideal CV cluster state.

\subsection{Equivalence of post-processing}\label{sec:hybridonebitGKP}

Since an ideal CV cluster state is a superposition of qubit cluster state in every displaced GKP basis, the many-qubit states downloaded to the ancillae will be a qubit cluster state. Without post-processing, the ancilla step after step 2 becomes
\begin{equation}
\prod_{k = 1}^N\qty[\hat{R}_{Z_k}(-\theta_k)] \prod_{i = 1}^N \qty[\hat{X}_i^{l_i}] \ket{G}.\label{eqn:onebitstate}
\end{equation}

Recall that the stabilizers of a qubit cluster state are the product of the $\hat{X}$ operator of any qubit and the $\hat{Z}$ operator of all its adjacent qubits, i.e. \cite{raussendorf_measurement-based_2003, briegel_persistent_2001, hein_entanglement_2006}, 
\begin{equation}
    \hat{X}_i \prod_{j \in n(i)}  \hat{Z}_j \ket{G} = \ket{G}. \label{eqn:stabilizer}
\end{equation}
We multiply all stabilizers together and move all $\hat{X}$ operators to the left of $\hat{Z}$ operators,
\begin{equation}
        \prod_{i = 1}^N \qty[\hat{X}^{l_i}_i \prod_{j \in n(i)}  \qty[\hat{Z}_j^{l_i}]] \ket{G}= \pm \prod_{k = 1}^N \qty[\hat{X}^{l_k}_k] \prod_{(i,j) \in E} \qty[\hat{Z}_i^{l_j}] \ket{G} =  \ket{G}. \label{eqn:multistabilizer}
\end{equation}
where the $\pm$ sign in the first identity arises from the anti-commutation relation of $\hat{X}$ and $\hat{Z}$, and the precise sign generally depends on the structure of graph. We have also used the identity $\prod_{(i, j) \in E}  \hat{Z}_j^{l_i} = \prod_{(i, j) \in E}  \hat{Z}_i^{l_j}$ and $l_j = \sum_{i} A_{ij} l_i$ to relabel the indices. Substituting \cref{eqn:multistabilizer} into \cref{eqn:onebitstate} we get
\begin{align}
    \begin{split}
    \prod_{k = 1}^N\qty[\hat{R}_{Z_k}(-\theta_k)] \prod_{i = 1}^N \qty[\hat{X}_i^{l_i}] \ket{G} 
    &= \pm \prod_{k = 1}^N \prod_{j \in n(k)} \qty[\hat{R}_{Z_k}(-\theta_k)  e^{i \pi l_j \hat{Z}_k} ]\ket{G} = \pm \prod_{k = 1}^N \qty[\hat{R}_{Z_k}(-\phi_k)] \ket{G}.\label{eqn:XvZpostprocessing}
    \end{split}
\end{align}
In the last step, we have used $\hat{R}_{Z_i} (A_{ij}(2m_i + l_i)\pi) = \hat{R}_{Z_i} (A_{ij}l_i\pi) = e^{i \pi l_j \hat{Z}_i}$, for all $j \in n(i)$ which implies $A_{ij} = 1$. The phase-shift required to correct the by-product in \cref{eqn:XvZpostprocessing} is exactly what we suggested in Eq. (1) in the main text. We note that the $\pm$ sign becomes an unimportant global phase.

\section{Proof of Equation 5} \label{chap:finsq}

Consider any general pure qumode state $\ket{\phi} = \int_{-\infty}^{\infty} dq \, \phi(q) \ket{q}_q $ being conditional-displaced by the qubit, the joint state is given by $\hat{C}_D \ket{\phi}\ket{+} = \frac{1}{\sqrt{2}}\int^{\infty}_{-\infty} dq \, \phi(q)\ket{q}\ket{0} + \phi(q-\sqrt{\pi})\ket{q}\ket{1}$. Then the qumode is measured in $q$ quadrature, the qubit state becomes
\begin{align}
    \begin{split}
    \ket{\Psi(q)} \equiv \frac{1}{\mathcal{N}_q} \qty(\phi(q) \ket*{0} + \phi(q-\sqrt{\pi})\ket*{1}) \propto \bra{q}_q \hat{C}_{D} \ket*{\phi} |+\rangle,
    \end{split}
    \label{eqn:psiq}
\end{align}
which depends on the measurement outcome $q$ and $\mathcal{N}_q = \sqrt{|\phi(q)|^2 + |\phi(q-\sqrt{\pi})|^2}$. 

Suppose the mode is initially in a $p$-displaced squeezed state $\ket{\psi} = \hat{D}_p(p_0) \hat{S}(r_0) \ket{\mathrm{vac}}$ with displacement $p_0$. Its wave function is given by \cite{leonhardt_essential_2010}
\begin{align}
    \psi(q; p_0) &= \frac{1}{\sqrt[4]{\pi e^{2r_0}}} \exp(-\frac{q^2}{2 e^{2r_0}} + i p_0 q), \label{eqn:displacedsqueezedwavefunction}
\end{align}
where the variance in $q$ is $e^{2r_0}/2$. After conditional displacement and $q$ measurement, the qubit is projected to
\begin{align}
    \begin{split}
        \ket*{\Psi(q; p_0)} &\equiv \frac{1}{\mathcal{N}_q} \qty(e^{-q^2/ (2 e^{2r_0})}\ket*{0} + e^{-i p_0 \sqrt{\pi}}e^{-(q-\sqrt{\pi})^2/ (2 e^{2r_0})}\ket*{1}), \label{eqn:qubitdispsq} \\
        &\propto \psi(q; 0) \ket{0} + e^{-ip_0 \sqrt{\pi}} \psi(q - \sqrt{\pi} ; 0) \ket{1},
    \end{split}
\end{align}
with normalization factor $\mathcal{N}_q = \sqrt{e^{-q^2/ (2 e^{2r_0})}+e^{-(q-\sqrt{\pi})^2/ (2 e^{2r_0})}}$. This yields Eq. (5) in the main text.

\section{Proof of Equation 6} \label{sec:weakmeasurecorrection}

Suppose we have a qubit $(\ket{0} + e^{i \alpha} \gamma \ket{1})/(\sqrt{1+\gamma^2})$, where the amplitude imbalance factor $\gamma$ is known but the relative phase $\alpha$ is unknown. The amplitude balancing POVM being applied depends on the known value of $\gamma$.

\textbf{Case 1. } If $\gamma < 1$, i.e. the qubit has a higher amplitude of $\ket{0}$ than $\ket{1}$. The POVMs takes the form
\begin{align}
    \hat{M}_0 = \gamma \ketbra{0}{0} + \ketbra{1}{1}, \quad \hat{M}_1 = \sqrt{1 - \gamma^2}\ketbra{0}{0}.\label{eqn:M0M1}
\end{align}
    The correction succeeds if $\hat{M}_0$ is obtained, which reduces the amplitude of $\ket{0}$ to balance with that of $\ket{1}$. The probabilities of getting each measurement outcomes are
\begin{equation}
    p_{m=0} = \frac{2\gamma^2}{1 + \gamma^2} , \quad p_{m=1} = \frac{1-\gamma^2}{1+\gamma^2},
\end{equation}
and the states after measurement are respectively
\begin{equation}
    \ket*{\Psi_{m=0}} = \frac{1}{\sqrt{2}}(\ket*{0} + e^{i\alpha}\ket*{1}), \quad \ket*{\Psi_{m=1}} = \ket*{0}.
\end{equation}

\textbf{Case 2.} If $\gamma > 1$, i.e. the qubit has a higher amplitude of $\ket{1}$ than $\ket{0}$, we consider the POVMs 
\begin{align}
    \hat{M}_0' = \ketbra{0}{0} + \gamma^{-1} \ketbra{1}{1}, \quad \hat{M}_1' = \sqrt{1 - \gamma^{-2}}\ketbra{1}{1}. \label{eqn:M0M1_2}
\end{align}
The correction succeeds if $\hat{M}_0'$ is obtained, which reduces the amplitude of $\ket{1}$ to balance with that of $\ket{0}$. In this case, the probabilities are
\begin{equation}
    p_{m'=0} = \frac{2}{1+\gamma^2}, \quad p_{m'=1} = \frac{\gamma^2-1}{\gamma^2+1},
\end{equation}
and the states after measurement are respectively
\begin{equation}
        \ket*{\Psi_{m'=0}} = \frac{1}{\sqrt{2}}(\ket*{0} + e^{i\alpha}\ket*{1}), \quad \ket*{\Psi_{m'=1}} = \ket*{1}.
\end{equation}

In our case that the amplitude imbalance is introduced through interacting with a finitely squeezed state, i.e. $\gamma = |\psi(q-\sqrt{\pi};0)|/ |\psi(q;0)|,$ the expected successful probability of correction can be calculated by integrating the probability of obtaining each value of $q$, multiplying with the successful correction (i.e. getting $m=0$ or $m'=0$) at each $q$: 
\begin{align}
    p_\mathrm{succ} &= \int^\infty_{-\infty} P(q) p_{m, m'=0} dq = \int^{\sqrt\pi/2}_{-\infty} |\psi(q-\sqrt{\pi};0)|^2 dq + \int_{\sqrt\pi/2}^{\infty} |\psi(q;0)|^2 dq, \label{eqn:success} 
\end{align}
where $P(q) \equiv \qty(|\psi(q;0)|^2 + |\psi(q-\sqrt\pi;0)|^2)/2$.

By substituting the wave function of squeezed state in \cref{eqn:displacedsqueezedwavefunction}, we get the analytic expression of the successful probability
\begin{equation}
p_\mathrm{succ} = \text{erfc}\left(\frac{\sqrt{\pi}e^{-r_0}}{2}\right). \label{eqn:successfail2}
\end{equation}
Since a failed correction results in a qubit erasure, we have $p_\mathrm{del} = 1 - p_\mathrm{succ}$ and hence Eq. (6) in main text.

\section{Multiple-qumode-per-vertex architecture}\label{sec:suppressing}

\begin{figure}[H]
    \centering
    \includegraphics[width=0.9\linewidth]{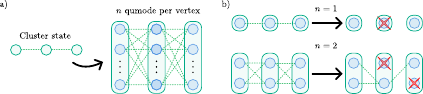}
    \caption[Multi-qubit-per-site-model.]{(a) Multiple-qumode-per-vertex architecture. Each vertex consists of $n$ qumodes that every of them is connected with all qumodes in the adjacent vertices. (b) A vertex is disconnected from the graph if it consists of only one qumode $(n=1)$ and the qubit is deleted.  However, deleting one qubit does not disconnect the vertex if it consists of two qumodes $(n=2)$.}
    \label{fig:MultiQubit}
\end{figure}

To prepare a qubit cluster state with graph $G$ in our current scheme, we prepare a CV cluster state with the same graph, which each vertex is associated with one qumode. However, because the CV cluster state is finitely squeezed, after entanglement downloading and weak measurement corrections, some qubits will be probabilistically deleted from $G$. 

In addition to increasing squeezing, we are able to further suppress the deletion probability $p$ by considering a multiple-qumode-per-vertex architecture, as shown in \cref{fig:MultiQubit}(a) \cite{azuma_all-photonic_2015,borregaard_one-way_2020}. The idea is to associate multiple qumodes to each vertex, and generate a cluster state that every qumode in a vertex is connected to all qumodes in the adjacent vertices.  We are still considering attaching each qumode a qubit to which the entanglement is downloaded.  After our protocol, we will end up with an imperfect qubit cluster state that each vertex consists of multiple, say $n$, qubits. It is imperfect because and the amplitude of each qubit is imbalance due to the finite squeezing of the qumodes. The amplitude balancing is then conducted, which qubits could be deleted probabilistically.  However, in this scheme deleting a qubit does not necessarily disconnect a vertex from the graph, because other qubits belonging to that vertex could still be connecting with the adjacent vertices. If the probability of a qubit deletion is $p$, then the probability of a vertex being disconnected from the graph is $p^n < p$, which is exponentially suppressed. Afterwards, a qubit cluster state can be formed by removing all but one connected qubit by computational basis measurement.

\cref{fig:MultiQubit}(b) illustrates the idea of suppressed vertex disconnection with the cases of $n=1$ (single qumode per vertex, the scheme discussed in main text) and $n=2$ (two qumodes per vertex).
\section{Proof of Equation 7} \label{chap:thermalstate}

Here, we study the single qubit error due to initial thermalization of the qumode. We first consider the relationship between squeezed thermal states and displaced squeezed states. The Wigner function of a squeezed thermal state is given by 
\begin{equation}
    W_{sq}(q, p; r, \bar{n}) = \frac{1}{(1+2\bar{n})\pi}\exp(-\frac{(e^{2r} p^2+e^{-2r}q^2)}{1+2\bar{n}}), \label{eqn:sqth}
\end{equation}
and that of a displaced squeezed state with displacement $(0, p_0)$ is given by 
\begin{equation}
    W_{ds}(q, p; r_0, p_0) = \frac{1}{\pi}\exp(-\qty(e^{2r_0}(p-\sqrt{2}p_0)^2 - e^{-2r_0}q^2)). \label{eqn:sqdp}
\end{equation}
It is then easy to show that a squeezed thermal state is a mixture of $p$-displaced squeezed state, i.e.
\begin{align}
    \begin{split}
    W_{sq}(q, p; r, \bar{n}) & = \int_{-\infty}^{\infty} W_{ds}(q, p; r_0, p_0) f(p_0, \sigma_{p_0}^2) dp_0 ,
    \end{split} 
    \label{eqn:intsqdp}
\end{align}
where $f(p_0, \sigma_{p_0}^2) \equiv (2\pi \sigma_{p_0}^2)^{-1/2} e^{-p_0^2/(2\sigma_{p_0}^2)}$ is a Gaussian distribution with mean $0$ and standard deviation $\sigma_{p_0}$. The squeezing strength $r$ and thermalization $\bar{n}$ of the squeezed thermal states are related to $\sigma_{p_0}$ and squeezing strength of the $p$-displaced squeezed state $r_0$ as
\begin{equation}
     e^{2r_0} = e^{2r}(1 + 2\bar{n}), \quad \sigma_{p_0}^2 = \frac{\bar{n}(1 + \bar{n})}{e^{2r}(1 + 2\bar{n})}. \label{eqn:conversion}
\end{equation}

Following \cref{eqn:qubitdispsq}, the density matrix of the resulting qubit after interaction with a squeezed displaced state is
\begin{align}
    \ketbra{\Psi(q; p_0)}{\Psi(q; p_0)} = \frac{1}{\mathcal{N}_q^2}\mqty[|\psi(q; p_0)|^2 & \psi(q; p_0)\psi(q-\sqrt{\pi}; p_0)e^{ip_0\sqrt{\pi}} \\ \psi(q;  p_0)\psi(q-\sqrt{\pi}; p_0)e^{-ip_0\sqrt{\pi}} & |\psi(q-\sqrt{\pi}; p_0)|^2]. 
\end{align}
We note that the normalization constant $\mathcal{N}_q$ is independent of the displacement $p_0$. To model the qubit after interacting with a squeezed thermal state, we integrate the qubit over the mixture of $p_0$ displacements,
\begin{align}
\begin{split}
    \rho_\Psi &= \frac{1}{\mathcal{N}_q^2} \int_{-\infty}^{\infty} dp_0 f(p_0,\sigma_{p_0}^2) \ketbra{\Psi(q ; p_0)}{\Psi(q ; p_0)} , \\
    &= \frac{1}{\mathcal{N}_q^2} \mqty[|\psi(q;0)|^2 & \psi(q;0)\psi(q-\sqrt{\pi};0)e^{-\frac{1}{2}a^2 \sigma_{p_0}^2} \\ \psi(q;0)\psi(q-\sqrt{\pi};0)e^{-\frac{1}{2}a^2 \sigma_{p_0}^2} & |\psi(q-\sqrt{\pi};0)|^2]. 
\end{split}
\end{align}
This resulting qubit state is equivalent to a qubit coupling to a zero-mean squeezed state and then undergoes dephasing, 
\begin{equation}
    \label{eqn:dephase2}
    \rho_{\Psi} = (1-p_\Psi)\ketbra{\Psi(q;0)}+ p_\Psi\hat{Z} \ketbra{\Psi(q;0)} \hat{Z},
\end{equation} 
where the dephasing rate is given by
\begin{equation}
    p_\Psi = \frac{1 - e^{-\frac{1}{2}\pi \sigma_{p_0}^2}}{2} \label{eqn:probdephase}.
\end{equation}
This yields Eq. (7) in the main text. 

\section{Uncorrelating imperfection noise} \label{chap:losschannel}

\subsection{Channel loss}
\begin{figure}[H]
    \centering
    \includegraphics[scale=1]{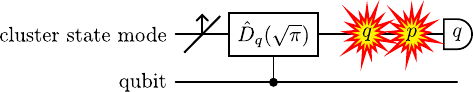}
    \caption{A single cluster state mode passing through a loss channel before interacting with qubit, and then followed by detector noise.}
    \label{fig:physicalpicture1}
\end{figure}

We consider each mode of a CV cluster state passes through a lossy channel with loss rate $\epsilon$ before interacting with the qubit. The channel can be modelled as a beam splitting interaction between the system and environment:
\begin{equation}
    \hat{a} \to  \sqrt{1-\epsilon}\hat{a} +  \sqrt{\epsilon} \hat{b}_1.
\end{equation}
where $\hat{a}$ is the system mode, and $\hat{b}_1$ is an environment mode.

\subsection{Detector inefficiency}

After interacting with the qubit, each mode is measured by a detector with efficiency $\epsilon_2$, which can be modeled as an ideal detector after a lossy channel:
\begin{equation}
    \hat{a} \to \sqrt{1-\epsilon_2}\hat{a}+\sqrt{\epsilon_2} \hat{b}_2.
\end{equation}
where $\hat{b}_2$ is an environment mode.

The measurement signal is then usually amplified to compensate the detection inefficiency. We model the signal amplification by a quantum-limited amplifier \cite{caves_quantum_1982}, so the combined effect of detection inefficiency and amplification is 
\begin{equation}
    \hat{a} \to \hat{a} + \sqrt{\frac{\epsilon_2}{1-\epsilon_2}}(b_2 + b_3^\dagger).
\end{equation}
where $\hat{b}_3$ is the environmental mode corresponding to amplification.

The combined effect is  equivalent to a random displacement in both quadratures \cite{noh_quantum_2019}. This can be explicitly seen if we define the collective modes
\begin{equation*}
    \hat{B}_2 = \frac{\hat{b}_2 + \hat{b}_3}{\sqrt{2}}, \quad \hat{B}_3 = \frac{\hat{b}_2 - \hat{b}_3}{\sqrt{2}},
\end{equation*}
where $\hat{Q}_i, \hat{P}_i$ are the quadratures of mode $\hat{B}_i$. Then the mode quadratures will be transformed as
\begin{align}
    \hat{q} \to \hat{q} + \sqrt{2}\qty(\sqrt{\frac{\epsilon_2}{1-\epsilon_2}} \hat{Q}_2 ) , \quad \hat{p} \to \hat{p} + \sqrt{2}\qty(\sqrt{\frac{\epsilon_2}{1-\epsilon_2}} \hat{P}_3).\label{eqn:detectorQP}
\end{align}
This shows that each quadrature of the system mode suffers from fluctuation noise that is induced by the quantum fluctuation of an independent environmental mode. We find that the quantum operation $\hat{U}_d$ corresponding to this noise process is given by

\begin{align}
    \hat{U}_d 
    &=\exp(ig_\mathrm{env}\hat{q} \hat{P}_3)\exp(-i\frac{g_\mathrm{env}^2}{2}\hat{Q}_2 \hat{P}_3)\exp(-ig_\mathrm{env}\hat{p} \hat{Q}_2), \label{eqn:BCHdetector0}
\end{align}
where $g_\mathrm{env} \equiv \sqrt{2 \epsilon_2}$.

However, not all operations in \cref{eqn:BCHdetector0} will introduce errors onto the qubit. Specifically, since the system mode will be measured in the $q$ quadrature,  $\exp(ig_\mathrm{env}\hat{q} \hat{P}_3)$ will become a displacement on mode $\hat{B}_3$. Furthermore, since both environmental modes $\hat{B}_2$ and $\hat{B}_3$ will be traced over, this displacement and the second term $\exp(-i\frac{g_\mathrm{env}^2}{2}\hat{Q}_2 \hat{P}_3)$ do not affect the system mode.

As a result, the only effective noise coming from detection inefficiency is the third term in \cref{eqn:BCHdetector0} $\exp(-i g_\mathrm{env} \hat{p} \hat{Q}_2)$, which is a random fluctuation of the system mode in $q$ quadrature, i.e.
\begin{multline}
\mathrm{tr}_{B_2 B_3}    \qty(\bra{q}_a\hat{U}_d \hat{C}_D \qty(\ketbra{\mathrm{vac}}{\mathrm{vac}}_{B_2 B_3} \otimes \rho_a\otimes  \ketbra{+}{+}) \hat{C}_D^\dagger \hat{U}_d^\dagger \ket{q}_a  ) \\ =  \bra{q}_a \hat{C}_D \qty( \mathrm{tr}_{B_2}\qty(
e^{-ig_\text{env}\hat{p}_a \hat{Q}_2} \qty(\ketbra{\mathrm{vac}}{\mathrm{vac}}_{B_2} \otimes \rho_a ) e^{ig_\text{env}\hat{p}_a \hat{Q}_2}) \otimes \ketbra{+}{+}  )  \hat{C}_D^\dagger \ket{q}_a ,\label{eqn:detectorloss1} 
\end{multline}
In the last relation, we commute the operator of random fluctuation with the conditional displacement, this is possible because both are along the $q$ quadrature. Therefore the $q$ quadrature noise can be lumped into the mode preparation error, as shown in Fig. 3(a) in the main text.

\subsection{Combining channel loss and detector inefficiency}

We now study the combined effects due to both channel loss and detection inefficiency. For any $N$-mode Gaussian initial state with covariance matrix $\mathbf{V}$, these errors will introduce the transformation
\begin{align}
    \begin{split}
    \mathbf{V} &\to \mathbf{T}_\mathrm{loss} \mathbf{V} \mathbf{T}_\mathrm{loss}^T + \mathbf{N}_\mathrm{loss} + \mathbf{N}_\mathrm{detector}, 
    \end{split}
    \label{eqn:physicalpicfinal}
\end{align}
where $\mathbf{T}_\mathrm{loss} = \sqrt{1-\epsilon} \mathbf{I}_{2n}, \mathbf{N}_\mathrm{loss} = \epsilon/2 \mathbf{I}_{2n}, \mathbf{N}_\mathrm{detector} = \text{diag}(g_\mathrm{env}^2/2 \mathbf{I}_n, 0)$ \cite{menicucci_graphical_2011}. This allows us to identify the imperfect effective many-mode state from which the qubits are downloading entanglement from. 

Suppose the input state is a thermal CV cluster state. We find that the resultant state in \cref{eqn:physicalpicfinal} is generally not a thermal CV cluster state. Therefore the downloaded error will not be single-qubit dephasing, but generally involves correlated noise between multiple qubits.

\subsection{Uncorrelating qubit noise by adding CV correlation}\label{sec:thermalCV=loss}

\begin{figure}[H]
    \centering
    \includegraphics[scale=1]{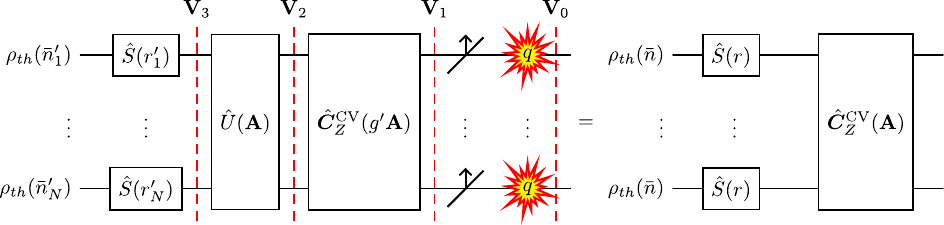}
    \caption[Elaborated version of Fig. 3b in the main text to suppress correlated noise on the qubits.]{Elaborated version of Fig. 3b in the main text which illustrates our strategy to suppress correlated noise on the qubits. After adding noise and using a beam splitter to correlate the modes, the state after loss and detector noise is equivalent a thermal cluster state, so that the qubit will experience uncorrelated error. We calculate the covariance matrices of the states in each stage from the back of the circuit. Here, $g' = \frac{B_1}{B_1 - C_1} \geq 1$.}
    \label{fig:loss-scheme1}
\end{figure}

Since the problem is the correlation that is generated by channel loss and detection inefficiency, our remedy is to include additional correlation into the cluster state, so that both correlations will cancel each other. In this way, the qubits will be effectively downloading entanglement from a thermal cluster state, thus the noise downloaded will be uncorrelated. 

Specifically, suppose we want the qubits to effectively interact with a thermal CV cluster state with covariance matrix $\mathbf{V}_0$, which has the following structure \cite{menicucci_graphical_2011}
\begin{equation}
    \mathbf{V}_0 = \mqty[B_1 \mathbf{I}_n & B_1 \mathbf{A} \\ B_1 \mathbf{A} & B_2 \mathbf{I}_n + B_1 \mathbf{A}{ \mathbf{A}}^T] =  \mqty[\mathbf{I}_n & 0 \\ \mathbf{A} & \mathbf{I}_n]\mqty[B_1 \mathbf{I}_n  & 0 \\ 0& B_2 \mathbf{I}_n] \mqty[\mathbf{I}_n & \mathbf{A} \\ 0 & \mathbf{I}_n], \label{eqn:generalcovCZ}
\end{equation}
where 
\begin{align}
    B_1&\equiv e^{2r}\qty(\bar{n} + \frac{1}{2}), \quad B_2\equiv e^{-2r}\qty(\bar{n} + \frac{1}{2}) ,\label{eqn:B1defn} 
\end{align}
and $e^{2r}$ and $\bar{n}$ are respectively the squeezing and thermal excitation of each mode. These parameters are to be determined from the amount of noise suffered by the system. 

Our task is to deduce the Gaussian state that should be prepared, so that it becomes a thermal CV cluster state after channel loss and detection inefficiency noise. Our approach is to trace back the covariance matrix at each stage of transformation, as shown in \cref{fig:loss-scheme1}, and impose conditions to guarantee that the state at each stage is physical. 

First, we look at the state just before channel loss and detection inefficiency, its covariance matrix $\mathbf{V}_1$ is related to $\mathbf{V}_0$ as
\begin{align}
    \mathbf{V}_0 
    &= \mathbf{V}_1+\mqty[C_1 \mathbf{I}_n & \mathbf{0} \\ \mathbf{0} & C_2 \mathbf{I}_n ]\label{deriv:sqloss4},    
\end{align}
with $C_1$ and $C_2$ are known parameters that are determined by the system imperfections:
\begin{align}
    C_1 = \frac{\epsilon}{2} + \frac{\epsilon_2}{1-\epsilon_2}, \quad C_2 = \frac{\epsilon}{2}, \label{eqn:C1C2defn}
\end{align}

Next, we find that the state with $\mathbf{V}_1$ can be constructed by applying an array of CPHASE gates onto a state that contains correlations only between like quadratures, for which the covariance matrix is given by $\mathbf{V}_2$, i.e. 
\begin{align}
    \begin{split}
    \mathbf{V}_1 &=  \mqty[\mathbf{I}_n & \mathbf{0} \\ g' \mathbf{A}  & \mathbf{I}_n]\underbrace{\frac{1}{1 - \epsilon} \mqty[(B_1 - C_1) \mathbf{I}_n & \mathbf{0} \\ \mathbf{0} & (B_2 - C_2) \mathbf{I}_n - \frac{B_1 C_1}{B_1 - C_1}{\mathbf{A}}^2]}_{\equiv \mathbf{V}_2}\mqty[\mathbf{I}_n & g' \mathbf{A} \\ \mathbf{0} & \mathbf{I}_n]. 
    \end{split}
    \label{eqn:reducedg2}
\end{align}
The array of CPHASE gates follows the same graph as the desired state, i.e. they share the same adjacency matrix $\mathbf{A}$, but with a magnified gate strength
\begin{equation}
    g' = \frac{B_1}{B_1 - C_1} \geq 1. \label{eqn:gstrength}
\end{equation}
We also note that the equivalent state needs to be generated by stronger CPHASE gates, i.e. \cref{eqn:gstrength}, because the imperfections reduce the entanglement among modes.

Finally, we discuss how to generate the state with covariance $\mathbf{V}_2$. Since $\mathbf{A}^2$ is real, symmetric, and positive semi-definite, it is orthogonally diagonalizable, i.e. $\mathbf{A}^2 = \mathbf{ODO}^T$ where $\mathbf{O}$ is some orthogonal matrix and $\mathbf{D}$ is a diagonal matrix with non-negative entries $D_{i} \geq 0$. These diagonal entries are also the eigenvalues of $\mathbf{A}^2$. For any orthogonal matrix $\mathbf{O}$, we can always construct a unitary sympletic matrix $\mathbf{U} = \text{diag}(\mathbf{O}, \mathbf{O})$ that is also orthogonal. An orthogonal symplectic matrix can represent a multi-mode transformation that is passively linear, which can always be implemented with a series of beam splitters \cite{weedbrook_gaussian_2012,reck_experimental_1994,clements_optimal_2016,bell_further_2021}.

Once we have identified the passive transformation $\mathbf{U}$, the covariance matrix $\mathbf{V}_2$ can be decomposed in terms of $\mathbf{U}$ and a diagonal matrix $\mathbf{V}_3$, 
\begin{align}
    \mathbf{V}_2 &= \mathbf{U}\mathbf{V}_3\mathbf{U}^T=\mqty[\mathbf{O} & 0 \\ 0 & \mathbf{O}] \underbrace{\frac{1}{1-\epsilon}\mqty[(B_1 - C_1) \mathbf{I}_n & 0 \\ 0 & 
    (B_2 - C_2) \mathbf{I}_n - \frac{B_1 C_1}{B_1 - C_1} \mathbf{D}]}_{\equiv \mathbf{V}_3} \mqty[\mathbf{O}^T & 0 \\ 0 & \mathbf{O}^T] \label{eqn:V2}. 
\end{align}

The matrix $\mathbf{V}_3$ does not always correspond to the covariance matrix of a physical state, because the variances in a physical covariance matrix should obey the Heisenberg uncertainty relation \cite{weedbrook_gaussian_2012}.  If it does, since $\mathbf{V}_3$ is diagonal, it corresponds to $N$ uncorrelated modes that each is a squeezed thermal state. We now impose the condition that $\mathbf{V}_3$ is physical, and evaluate the squeezing and thermalization of every mode. Since the upper left block of $\mathbf{V}_3$ is proportional to the identity, every mode has the same $q$ variance $(B_1-C_1)/(1-\epsilon)$. However, the bottom right block is diagonal but not proportional to the identity, so every mode has a different $p$ variance. We see that the mode with the smallest variance in $p$ corresponds to the one with the largest eigenvalue of $\mathbf{D}$, which we label as $D_{max}$. Let us highlight this mode as the principal mode.

We consider the most general $N$-mode zero-mean uncorrelated state, which can be characterized by the thermalization and squeezing strength of each mode are $\{\bar{n}_i'\}_{i=1}^N$, and $\{r_i'\}_{i=1}^N$ respectively. If $\mathbf{V}_3$ is physical, then it must have the expression
\begin{equation}
    \mathbf{V}_3 = \mqty[\oplus_{i=1}^{N} e^{2r_i'}(\bar{n}_i' + \frac{1}{2}) & 0 \\ 0 & \oplus_{i=1}^{N} e^{-2r_i'}(\bar{n}_i' + \frac{1}{2}) ]. \label{eqn:v3-1}
\end{equation}
By equating $\mathbf{V}_3$ in \cref{eqn:V2,eqn:v3-1} we have
\begin{align}
    \begin{split}
            e^{2r_i'} &= 
   \frac{B_1-C_1}{\sqrt{(B_1-C_1)
   (B_2-C_2)-B_1 C_1 D_i}}, \\
   \bar{n}_i' &= 
   \frac{\sqrt{(B_1-C_1) (B_2-C_2)-B_1
   C_1 D_i}}{1 - \epsilon }-\frac{1}{2}.
    \end{split}
   \label{eqn:v3-3}.
\end{align}

The physical requirement that $\exp(2 r_i') \geq 0$ and $\bar{n_i}\geq0$ imposes the following conditions on the system parameters
\begin{align}
    \begin{split}
    B_1 - C_1 &> 0, \\
    B_2 - C_2 - \frac{B_1 C_1}{B_1 - C_1} D_i &> 0, \\
    \frac{1}{(1 - \epsilon)^2}\qty(B_1 - C_1)\qty(B_2 - C_2 - \frac{B_1 C_1}{B_1 - C_1} D_i) &\geq \frac{1}{4}.
    \end{split}
    \label{eqn:physical1}
\end{align}

It is easy to verify that once the principal mode satisfies these conditions, i.e. \cref{eqn:physical1} are valid when $D_i = D_{max}$, then all other modes which $D_i \leq D_{max}$ will also satisfy these conditions. It means that if the principal mode is physical, then all other modes must be physical as well. Intuitively, it is because $D_i \leq D_{max}$ implies that the $p$ variance of all other modes is larger than that of the principal mode; on the other hand, all modes have the same $q$ variance, so all modes will satisfy the uncertainty principle. 

In principle, there can be a wide range of parameters $B_1$ and $B_2$ for which $\mathbf{V}_3$ is physical.  In practice, however, we observe that some values of $B_1$ and $B_2$ can be more desirable. Specifically, we want the modes to include as little noise as possible. Besides, generating higher squeezing is usually more technically challenging, so an implementation with lower squeezing is more favorable. Here we observe from \cref{eqn:v3-3} that the principal mode has both the highest squeezing $r_{prin}'$ and lowest thermalization $\bar{n}_{prin}'$ among all modes, i.e. $r_i' \leq r_{prin}'$ and $\bar{n}_i' \geq \bar{n}_{prin}'$.  To reduce the squeezing and thermal excitation among modes, we therefore consider the principal mode to contain no thermalization $\bar{n}_{prin}' = 0$, and the highest achievable squeezing $r_{prin}' = r'$. Then we can determine the parameters $B_1$ and $B_2$ from \cref{eqn:v3-3} as 
\begin{align}
\begin{split}
    B_1 &= C_1 + (1 - \epsilon)\frac{e^{2r'}}{2}, \\ 
    B_2 &= C_2 + C_1 D_{max} + \frac{2 C_1^2  D_{max}e^{-2r'}}{1 - \epsilon} + \frac{(1-\epsilon) e^{-2r'}}{2},
\end{split}\label{eqn:B1B2}
\end{align}
We can now use the definition in \cref{eqn:B1defn} with \cref{eqn:B1B2} to find the effective squeezing parameter and thermalization of the thermal CV cluster state that the qubits are interacting with: 
\begin{align}
    e^{2r} = \sqrt{\frac{B_1}{B_2}}, \quad \bar{n} = \sqrt{B_1 B_2} - \frac{1}{2}.  \label{eqn:nx}
\end{align}
Substituting \cref{eqn:B1B2} back into \cref{eqn:v3-3}, we can also find the squeezing and thermalization that should be prepared in all modes
\begin{align}
    \begin{split}
    e^{2r_i'} &= \frac{(1 - \epsilon) e^{2r'}}{\sqrt{4 C_1^2 \Delta_i +2 C_1 \Delta_i (1 - \epsilon) e^{2r'}  + (1 - \epsilon)^2}},  \\
    \bar{n}_i'  &= \frac{1}{2} \qty(\frac{\sqrt{4 C_1^2 \Delta_i+2 C_1 \Delta_i (1 - \epsilon) e^{2r'}  +(1 - \epsilon) ^2}}{(1 - \epsilon) }-1),
    \end{split}
    \label{eqn:rn}
\end{align}
where $\Delta_i \equiv D_{max} - D_i \geq 0$. 

Together with the equivalent circuit model, i.e. Eq. (4) in the main text, \cref{eqn:rn,eqn:nx} can be used to evaluate the influence of channel loss and detection efficiency on the downloaded qubit entanglement under our method. However, the nonlinear relation may be too complicated to analyze. To get some intuition about the detrimental effects, we consider the regime of weak errors, i.e. we expand \cref{eqn:gstrength,eqn:nx} to the linear order in channel loss $\epsilon$ and detection inefficiency $\epsilon_2$:
\begin{align}
    \begin{split}
    e^{2r} &\approx e^{2r'} - \epsilon \mathcal{W}_- -\epsilon_2 \mathcal{V}_-,\\
    \bar{n}  &\approx  \frac{\epsilon}{2}\qty( \mathcal{W}_+ e^{-2r'} - 1)+ \frac{\epsilon_2}{2}\qty(\mathcal{V}_+ e^{-2r'} ),\\
    \quad g' &\approx  1 + e^{-2r'} (\epsilon + 2\epsilon_2),
    \end{split}
\label{eqn:gthermalloss}
\end{align}
where $\mathcal{W}_{\pm} = (D_{max} e^{4r'} + e^{4r'} \pm 1)/2 \geq 0$ and 
$\mathcal{V}_{\pm} = \qty(D_{max} e^{4r'} \pm 1) \geq 0$. Because $\mathcal{W}_{\pm}, \mathcal{V}_{\pm} \geq 0$, we see that loss and detection inefficiency will reduce the squeezing and increase the thermalization of the equivalent cluster state, as expected. We note that this method is just one possible way to uncorrelate the noise in the downloaded qubit, but it does not mean to be the optimal way.  We expect there are other constructions that can further suppress the the downloaded qubit noise, for example, by allowing the effective cluster state to have different noise in each mode, i.e. the diagonalized matrix in \cref{eqn:generalcovCZ} to have more than two distinct eigenvalues.

\section{Weak conditional displacement} \label{chap:weakcoupling}

Here we show that the entanglement downloaded from $R$ identical copies of CV cluster state through weak interaction is the same as that downloaded from one copy of CV cluster state through a strong interaction. Consider qubit $i$ is interacting with the $i$-th modes of $R$ copies of a thermal CV cluster state, but the strength of conditional displacement is reduced to $\sqrt{\pi/R}$. The total CD operation applied on the qubit is given by

\begin{equation}
     \hat{I} \otimes \ketbra{0}{0} + \exp(-i \sqrt{\pi/R} \sum_{s=1}^R \hat{p}_i^{(s)} ) \otimes \ketbra{1}{1} = \hat{I} \otimes \ketbra{0}{0} + \exp(-i \sqrt{\pi} \sum_{s=1}^R \hat{P}_i ) \otimes \ketbra{1}{1}.
    \label{deriv:weakcoupled1}
\end{equation}

The last relation shows that we can treat qubit $i$ as effectively coupling to a collective mode $\hat{A}_i \equiv 1/\sqrt{R} \sum_{s=1}^R \hat{a}_i^{(s)}$, which the quadratures are
\begin{equation}
    \hat{Q_i} \equiv \frac{1}{\sqrt{R}}\sum_{s=1}^R \hat{q}_i^{(s)}, \quad \hat{P_i} \equiv \frac{1}{\sqrt{R}}\sum_{s=1}^R \hat{p}_i^{(s)}.
\end{equation}

What we need to show is that the collective cluster state, which  consists of the collective modes $(\hat{A}_1, \ldots \hat{A}_N)$, is the same as each copy of cluster state, which the $s$-th copy consists of the modes $(\hat{a}_1^{(s)}, \ldots \hat{a}_N^{(s)})$. It can be done by showing that they have the same covariance matrix, which the elements are given by
\begin{align}
    \begin{split}
        \expval{\hat{Q}_i \hat{Q}_j} &= \frac{1}{R}\expval{\sum_{s,s' = 1}^{R}\hat{q}_i^{(s)} \hat{q}_j^{(s')}} 
        = \frac{1}{R}\expval{\sum_{s=1}^{R}\hat{q}_i^{(s)} \hat{q}_j^{(s)}} = \expval{\hat{q}_i \hat{q}_j},\\
        \expval{\hat{P}_i \hat{P}_j} &= \frac{1}{R}\expval{\sum_{s,s' = 1}^{R}\hat{p}_i^{(s)} \hat{p}_j^{(s')}} 
        = \frac{1}{R}\expval{\sum_{s=1}^{R}\hat{p}_i^{(s)} \hat{p}_j^{(s)}} = \expval{\hat{p}_i \hat{p}_j},\\
        \expval{\hat{Q}_i \hat{P}_j + \hat{P}_j \hat{Q}_i} &= \frac{1}{R}\expval{\sum_{s,s' = 1}^{R}\hat{q}_i^{(s)} \hat{p}_j^{(s')} + \hat{p}_j^{(s')} \hat{q}_i^{(s')}}
        = \frac{1}{R}\expval{\sum_{s = 1}^{R}\hat{q}_i^{(s)} \hat{p}_j^{(s)} + \hat{p}_j^{(s)} \hat{q}_i^{(s)}} = \expval{\hat{q}_i \hat{p}_j + \hat{p}_j \hat{q}_i}.
    \end{split}
\end{align}
We have used the fact that the modes of copies $s$ and $s'$ are independent and identical, and all modes in a CV cluster state have zero mean in quadratures. This shows that in the weak interaction protocol the qubits are effectively downloading entanglement from a collective cluster state, which is identical to the cluster state in the strong measurement protocol. As a result, the errors downloaded in both cases are the same.

\end{document}